\documentclass[10pt,english]{article}
\usepackage{subfigure}
\usepackage{graphicx}
\usepackage{float}
\usepackage[T1]{fontenc}
\usepackage[latin9]{inputenc}
\usepackage[margin=1.in]{geometry}
\usepackage{float}
\usepackage{amsmath}
\usepackage{amsbsy}

\newtheorem{definition}{Definition}
\usepackage{setspace}
\usepackage{amssymb}
\usepackage{esint}
\newfloat{algorithm}{tbp}{loa}
\floatname{algorithm}{Algorithm}
\usepackage{algorithmic}
\usepackage{hyperref}
\usepackage{cite}

\newlength\myindent
\makeatletter

\floatstyle{ruled}
\newfloat{algorithm}{tbp}{loa}
\floatname{algorithm}{Algorithm}

\usepackage{babel}

\begin{document}

\title{Secret sharing scheme based on hashing}

\author{M. Andrecut}


\maketitle
{

\centering Calgary, Alberta, Canada

\centering mircea.andrecut@gmail.com

} 

\bigskip 

\begin{abstract}

We propose an adaptive threshold multi secret sharing scheme based solely on cryptographically secure hash functions. 
We show that the proposed scheme is also: perfect, ideal, verifiable, and proactive. 
Moreover the proposed scheme has a low computational complexity comparing to the most common schemes 
operating over finite fields.

\smallskip


\end{abstract}

\bigskip

\section{Introduction}

A secret sharing scheme (SSS) is a method for distributing a secret $S$, such as an encryption key, among a group of $n$ participants \cite{key-1}. 
In a SSS, a dealer $D$ allocates a share $s_i$ of the secret to each participant, $i=0,1,...,n-1$, such that the secret can only 
be reconstructed when the shares are combined together, and therefore the individual shares are of no use on their own \cite{key-2}. 
SSSs are used for storing and providing access to highly sensitive and important information, with applications in: cloud computing, military, intelligence, banking, health care, sensor networks etc. 

The additive SSS is probably the most simple method, and assumes that the secret $S$ can be split into randomly picked shares that are members of an Abelian group $s_i \in (G,+)$, $i=0,1,...,n-1$: 
$S = \sum_{i=0}^{n-1}s_i$. In this scheme, the secret can only be reconstructed by adding all the shares, since any subset of shares reveals nothing about $S$. 
An important extension is the threshold-SSS, or the $(t,n)$ SSS, where a threshold number $t$ of shares, less than the total number of shares $n$, $t\leq n$, is required to reconstruct the secret. 
This increases the robustness of the SSS, and avoids the failure to recover the secret in case one or few participants are unavailable. 

The first $(t,n)$ SSSs were created independently by Adi Shamir \cite{key-3} and George Blakley \cite{key-4} in 1979. 
For example, Shamir's scheme is based on polynomial interpolation, and the observation that $t+1$ different points are necessary to exactly define a $t$ order polynomial.
The secret $S$ corresponds to the first coefficient of the polynomial, and the remaining coefficients are set randomly. 
The values of the polynomial at different points $x_i$, $i=0,1,...,n-1$, are then distributed among $n$ participants as shares. 
The polynomial can be then reconstructed using Lagrange's interpolation method on $t$ or more shares, and therefore one can recover the secret corresponding to the first coefficient. 

Since their inception, many other SSSs have been proposed \cite{key-5}. However, most of them are infeasible in practical applications, and suffer more or less from the following problems:
\begin{itemize}
\item Lack of practical verification methods, requiring honesty from the dealer and/or the participants. 
\item Only one secret is typically shared, without the ability to share multiple secrets.
\item Threshold or secret modification requires a complete recalculation and redistribution of the shares.
\item High computational complexity, since operations are performed over Galois finite fields $GF(p)$ where $p$ is a large prime number. 
\item Some generalized schemes require an exponentially large number of shares \cite{key-6}. 
\end{itemize}
Here we propose an adaptive threshold SSS, based solely on cryptographically secure hash functions, and we show that this approach can successfully address the above problems.

\section{Preliminaries}

\subsection{One-way functions}

\begin{definition}{(One-way function)}
A one-way function is easy to compute on every input, but extremely hard to invert. 
\end{definition}
Thus, one can easily calculate $f(x)$ for any given $x$, but it is practically "impossible" to calculate $x$ given $f(x)$. While the existence of one-way functions 
is not mathematically proven, they are frequently used as an abstraction element in cryptography, and it can be shown that they are very well approximated by 
cryptographically secure hash functions \cite{key-2}.

\subsection{Hash functions}

A one-way cryptographic hash function is designed to transform input messages of any length into fixed-length output hash values, and it 
is can be defined as following \cite{key-7}:

\begin{definition}{(Hash Function)}
A function $h:\lbrace 0,1 \rbrace^* \rightarrow \lbrace 0,1 \rbrace^\ell$ which takes a bitstring $x$ of
arbitrary finite length (called message), and outputs a bitstring $h(x)$ (called hash) of fixed length $\ell$, is a hash function if it satisfies the following properties:

\begin{enumerate}

\item Ease of computation: it computes a hash value for any given bitstring; 

\item Deterministic computation: the hash function gives the same result for the same input data, any change in the input bitstring triggers a change in the hash value; 

\item Preimage resistance: it is impossible to invert, i.e. to generate a bitstring that has a given hash value;

\item Second preimage resistance: for a given bitstring and its hash it is impossible to find a different bitstring having the same hash value; 

\item Collision resistance: it is impossible to find two different bitstring messages with the same hash value. 

\end{enumerate}
\end{definition}

The above definition describes an "ideal" hash function. In reality, all the existing hash functions suffer from the generic birthday paradox attack, which means that 
for an $\ell$ bit hash function and $2^{\ell/2}$ different messages there exists a collision with no negligible probability. Therefore, the larger the bit length of the 
hash function the better collision resistance it has. Regarding the preimage and second preimage, currently the best generic attack to find a preimage is to perform 
exhaustive search on the sets of all possible messages (which in this case are hash values of length $\ell$), with a cost of $2^\ell$.

Typical hash functions are the standard SHA-256 and SHA-512 cryptographic hash functions, published by NIST \cite{key-6}. 
For example using the SHA-512 function, no matter how big the input data is, the output will always have a fixed 512-bits length. 
This property becomes critical when dealing with large data and transactions, because instead of remembering the input data, one can just remember the hash value. 
Thus, using a cryptographic hash function one can easily verify that a given data maps to a given hash value, but if the input data is unknown it is "impossible" 
to reconstruct the data (or any equivalent alternatives) by only knowing the hash value. 

These properties of the hash function are used to guarantee the security of the proposed SSS. 
Here, unless otherwise stated, we consider cryptographically secure hash functions with a length of at least 512 bits, such as SHA-512 for example.

\subsection{Access structure}

In a $(t,n)$ SSS we assume that there are $n$ participants $P=\{P_0,P_1,...,P_{n-1}\}$, and a minimum number of $t$ participants is required to recover the secret $S$. 

\begin{definition}{(Authorized Subset)}
An authorized subset is a group of participants that can recover the secret when they join their shares together. 
Reciprocally, an unauthorized subset is any group of participants that cannot recover the secret.
\end{definition}

\begin{definition}{(Access Structure)}
The set of all authorized subsets defines the access structure $Q$.
\end{definition}

\begin{definition}{(Minimal Authorized Subset)}
$A \in Q$ is a minimal authorized subset if for all $B \subseteq A$ we have $B \notin Q$.
\end{definition}

\begin{definition}{(Access Structure Basis)}
The basis $\tilde{Q}$ of an access structure $Q$ consists of all minimal authorized subsets.
\end{definition}

\subsection{Perfect and ideal SSS}

\begin{definition}{(Perfect SSS)}
A SSS that does not allow partial information about the shares to be disclosed. 
\end{definition}

\begin{definition}{(Ideal SSS)}
A SSS where the shares and the secret come from the same domain and have the same size.
\end{definition}

\subsection{Verifiable SSS}

\begin{definition}{(Verifiable SSS)}
A SSS is verifiable if it provides a method to check the validity of the shares.
\end{definition}
A verifiable SSS should be able to detect potential maliciousness from both the dealer and the participants. This means that the 
participants should be able to verify the validity of their shares, and reject potentially false shares provided 
by a malicious dealer. Reciprocally, at the secret recovery phase, the dealer should also be able to verify the shares provided by potentially 
malicious participants.

\subsection{Proactive SSS}

\begin{definition}{(Proactive SSS)}
A SSS is proactive if it is able to reset and redistribute new shares periodically, with or without the need of changing the secret. 
\end{definition}
The goal of a proactive SSS is to render useless any information gathered by an adversary between successive resets of the shares. 

\section{The proposed SSS}

\subsection{Main problem}

The main elements of the SSS are:

\begin{itemize}
\item $n>0$ participants: $P=\{P_0,P_1,...,P_{n-1}\}$.
\item $m\geq 1$ secrets: $S=\{S^0,S^1,...S^{m-1}\}$. 
\item the dealer $D$ (a trusted entity), who prepares and distributes the shares to the participants.
\item the cryptographically secure hash function $h()$, which is publicly known to everybody.
\end{itemize}
The problem we are seeking to solve is to build a perfect, ideal, verifiable and proactive SSS, 
that provides a secure access structure to the $m$ secrets from the group of $n$ participants, with a potentially modifiable and adaptable threshold $0<t\leq n$. 

\subsection{Encoding phase}

Here we assume that the $m$ secrets are stored in $m$ secure "vaults", and in fact only the secret access keys ("passwords") to the vaults must be shared among the participants. 
Another equivalent option would be to encrypt the secrets using a symmetric algorithm (AES for example), using these secret keys. 

We denote the secret keys by: $k_j$, $j=0,1,...,m-1$. 
Thus, we need to devise a method able to generate the keys using $t$ distinct secret shares. 
First we notice that each secret $S^j$ can be easily identified by associating it with its own hash value:
\begin{equation}
j\longleftrightarrow q_j = h(S^j), \quad j=0,1,...,m-1.
\end{equation}
Let us assume that the shares "basis" consists of a set of $t+1$ randomly generated bit-strings of length $\ell'$: 
\begin{equation}
s^* = \{ s^*_b \vert s^*_b \in \{0,1 \}^{\ell'}, b=0,1,...,t \}.
\end{equation}
Any two shares can be ordered using the operator $<$, and we say that a tuple $(s^*_i, s^*_j)$  is ordered if $s^*_i < s^*_j$. Subsequently, 
any list of distinct shares can be ordered using the $\varphi()$ function. An ordered list of shares can be concatenated 
into a string using the $\gamma()$ function. 
Here, we use $||$ as the concatenation operator between two strings, $a||b=ab$, and $\gamma()$ as a concatenation function 
applied to a list, $\gamma([a,b,c])=abc$.

To increase flexibility, we also assume that the dealer is one of the participants, and it has allocated its own share from the basis set $s^*$, $s^*_0$ for example. 
The rest of the participants are divided into $t$ groups $\Gamma_b$, $b=1,...,t$, where each group 
contains at least one member $n_b \geq 1$, such that: $n = \sum_{b=1}^t n_b$. 
Each member of such a group $\Gamma_b$ receives the corresponding distinct share $s^*_b$ from the basis set $s^*$. 

The secret keys generation and reconstruction will require that at least one member of each group participates in the process. This 
requirement also defines an authorized subset. One can see that in total there are $\prod_{b=1}^{t} n_b$ minimal authorized subsets. 
Therefore any authorized subset of participants is in fact required to be in the possession of the basis set $s^*$ in order to be able to reconstruct the secret keys. 

Let us denote by:
\begin{equation}
\xi = \gamma(\varphi(s^*)) 
\end{equation}
the string resulted by concatenating the ordered list of shares from the basis set. 
Each secret key $k_j$ is defined as the hash value of the concatenation of the hash value $q_j$ of the secret $S^j$ with the 
string $\xi$ resulted by concatenating the ordered basis set $s^*$, such that one can identify which key is about to be constructed or reconstructed:
\begin{equation}
j \longleftrightarrow q_j\longleftrightarrow h(q_j||\xi) = k_j, \quad j=0,1,...,m-1. 
\end{equation}

It is important to note that if the length $\ell'$ of the shares $s^*_b$ and the length $\ell$ of the secret keys $k_j$ is the same, $\ell'=\ell$, then the SSS becomes perfect and ideal. 
It is also important to note that the dealer can be "neutral" in this scheme by simply setting its share to an empty string such that it doesn't affect the computations of the secret keys. 

\subsection{Distribution phase}

Once the shares are generated and the secret keys are computed, the shares can be distributed to the participants. 
Due to the one-wayness property of the hash function one cannot learn anything about the shares from their hash values. 
Therefore, the dealer can also compute the hash value of each share, $g_b = h(s^*_b)$ from the basis set, and then publish the set of the resulted hashes: 
\begin{equation}
g^* = \{g_b \vert g_b = h(s^*_b), b = 0,1,...,t \}.
\end{equation}
Similarly, one cannot learn anything about the secrets from their hash values. Therefore, the dealer could also publish the hash values of the secrets. 
However, since a secret key $k_j$ also depends on the secret hash value $q_j=h(S^j)$, we prefer to avoid disclosing this hash value since it provides partial key information, and instead the dealer should disclose the 
second hash value $r_j=h(h(S^j))$:
\begin{equation}
r^* = \{r_j \vert r_j = h(h(S^j)), j = 0,1,...,m-1 \},
\end{equation}
which does not provide useful information about the secret $S^j$ and its first hash value $h(S^j)$.

The purpose of this publishing step is to make the SSS verifiable. This way any participant can verify if it has received a valid share from the dealer, by simply 
checking if the hash of the share $h(s_i)$ is included into the published set $g^*$, or not. We can therefore call $g^*$ as the "shares verification set". 
Also, any participant can verify if the second hash value of the recovered secret is included in the "secrets verification set" $r^*$, and therefore decide if it is a valid secret, or not. 

\subsection{Recovery phase}

Any authorized subset of participants can recover the secret $S^j$ by submitting their shares to the "combiner", and the index $j$ of the required secret. 
We should note here that the combiner may be the dealer, or a different trusted entity, the SSS will work the same in both cases. For simplicity purposes we assume that 
both the dealer and the combiner are controlled by the same trusted entity. 
The combiner can easily check if the submitted shares are valid by simply checking if their hash is included in the published verification set $g^*$. 
The combiner then extracts the set of distinct shares from the list of received shares. The number of distinct shares should be equal to the threshold value $t$. 
If this is true, the combiner has received the required set of basis shares, $s^*$, and it can compute the secret key $k_j=h(q_j||\gamma(\varphi(s^*)))$. 
The recovered secret $S^j$ is then shared with the members of the authorized subset, who then can verify 
that the secret is a valid one by checking if $h(h(S^j))$ is in $r^*$, or not.

\subsection{Proactive phase}

In order to prevent the adversary from learning anything about the shares, the dealer can periodically generate new shares and distribute them to the participants. 
We should note that the modification of the secrets $S^j$ does not require a recalculation of the shares, only the secret keys $k_j$ need a recalculation, since they depend on the 
hash value of the secret $q_j=h(S^j)$. 
Such an update can be also used to modify the threshold value if necessary. 

The dealer can also take an immediate protection step by simply refreshing just its own basis share, or just a few of the basis shares, and notify just the affected participants. 
Also, the dealer can revoke a given basis share, if for example a given group is not supposed to have access anymore. Revocation will obviously decrease the threshold by one unit, 
but if initially the threshold was relatively high the effect will be minimal on the SSS. 

A simple way to maintain a reasonable threshold value is to allocate more distinct shares to the dealer, 
which will also be considered basis shares. We call these as "controlling" shares. Another "adaptive" way to maintain the same threshold value is to replace the revoked share with a controlling share.
Similarly, when adding a new ordinary basis share, the dealer can revoke and remove a "controlling" share. This simple adaptive mechanism of creation and annihilation of controlling shares 
can therefore be used to maintain (conserve) a desired constant threshold value for the SSS.

\subsection{Observation}

In order to reduce the risk of disclosing shares to adversaries, the communication between the dealer (combiner) and the participants can be done by using any standard public key 
encryption method, such as RSA or ECC \cite{key-1}, \cite{key-2}. This way the shares can be encrypted using the public keys and decrypted using the private keys of the parties involved. 

While the proposed SSS is a $(t,n)$ threshold scheme, it operates differently than the Shamir's scheme. 
In Shamir's scheme, any $t$ participants can recover the secret, while in the proposed SSS the secrets can be recovered only by authorized subsets of participants. 
However, the minimal authorized set does require a minimum of $t$ participants to be in the possession of the basis shares. 

\section{Conclusion}

We have presented a threshold adaptive multi secret sharing scheme based solely on cryptographically secure hash functions. 
Additionally, we have shown that the proposed scheme is also: perfect, ideal, verifiable, and proactive. 
The proposed scheme also has a very low computational complexity and storage requirements, comparing to the most common schemes operating over finite fields.

\end{document}